\PassOptionsToPackage{numbers,sort & compress}{natbib}
\documentclass[11pt,showpacs,showkeys]{revtex4}
\usepackage{bbold}
\usepackage{float}
\usepackage[pagebackref=false,colorlinks,linkcolor=blue,citecolor=magenta]{hyperref}
\usepackage{subfigure}
\usepackage{amssymb}
\usepackage{amsmath}
\usepackage{amsfonts}
\usepackage{txfonts}
\usepackage{fontenc}
\usepackage{tone}
\DeclareMathAlphabet{\mathcal}{OMS}{zplm}{m}{n}
\usepackage{euscript}
\usepackage{bm} 
\usepackage{graphicx}
\usepackage{color}
\usepackage{multirow}

\begin{document}
	\title{Superintegrability on the Dunkl oscillator model in three-Dimensional Spaces of Constant Curvature}
	\author{Shi-Hai Dong $^{1,2}$}
	\email[ ]{dongsh2@yahoo.com}
\author{Amene Najafizade$^{3}$}
\email[Corresponding auther: ]{Najafizade1816@gmail.com}
\author{Hossein Panahi$^{3}$}
\email[ ]{t-panahi@guilan.ac.ir}
\author{Won Sang Chung$^{4}$}
\email[ ]{mimip44@naver.com}
\author{Hassan Hassanabadi$^{5}$}
\email[ ]{h.hasanabadi@shahroodut.ac.ir}
	\affiliation{$^{1}$Huzhou University, Huzhou, 313000, P. R. China}
	\affiliation{$^{2}$Laboratorio de Informaci\'{o}n Cu\'{a}ntica, CIDETEC, Instituto Polit\'{e}cnico Nacional, UPALM, CDMX07700, Mexico}
\affiliation{$^{3}$Department of Physics, University of Guilan, Rasht 51335-1914, Iran }
\affiliation{$^{4}$Department of Physics and Research Institute of Natural Science,
	College of Natural Science, Gyeongsang National University, Jinju 660-701, Korea}
\affiliation{$^{5}$Faculty of Physics, Shahrood University of Technology, Shahrood, Iran}
\date{\today}
\begin{abstract}
This paper has studied the three-dimensional Dunkl oscillator models in a generalization of superintegrable Euclidean Hamiltonian systems to curved ones. These models are defined based on curved Hamiltonians, which depend on a deformation parameter of underlying space and involve reflection operators. Their symmetries are obtained by the Jordan-Schwinger representations in the family of the Cayley-Klein orthogonal algebras using the creation and annihilation operators of the dynamical $sl_{-1}(2)$ algebra of the one-dimensional Dunkl oscillator. The resulting algebra is a deformation of $so_{\kappa_1\kappa_2}(4)$ with reflections, which is known as the Jordan-Schwinger-Dunkl algebra $jsd_{\kappa_1\kappa_2}(4)$. Hence, this model is shown to be maximal superintegrable. On the other hand, the superintegrability of the three-dimensional Dunkl oscillator model is studied from the factorization approach viewpoint. The spectrum of this system is derived through the separation of variables in geodesic polar coordinates, and the resulting eigenfunctions are algebraically given in terms of Jacobi polynomials. 
\end{abstract}
\maketitle
\section{Introduction}
 This paper is concerned with a new superintegrable generalisation of the three-dimensional isotropic Dunkl oscillator Hamiltonian in the flat plane involving reflection operators \cite{Hamiltonian}
 \begin{equation}\label{eq1}
 H=-\frac{1}{2}\left[(D_x^{\mu_x})^2+(D_y^{\mu_y})^2+(D_z^{\mu_z})^2\right]+\frac{\omega^2}{2}\left( x^2+ y^2+ z^2\right),
 \end{equation}
 to the curved spaces, where the frequency $\omega$ is an arbitrary real number, and 
 $D_{x_i}^{\mu_i}(x_i=x,y,z)$ is introduced as the Dunkl derivative \cite{derivative}
 \begin{equation}
 D_x^{\mu_i}=\partial_{x_i}+\frac{\mu_i}{x_i}(1-R_i), 
 \end{equation}
 Also, the square of the Dunkl derivative is in the following explicit form
 \begin{equation}\label{de}
 (D_{x_i}^{\mu_i})^2=\partial_{x_i}^2+2\frac{\mu_i}{x_i}\partial_{x_i}-\frac{\mu_i}{x_i^2}(1-R_i).
 \end{equation}
In the meantime, $R_i$ plays the role of the reflection operator concerning the $x_i=0$ axis and it has the properties of  
 \begin{equation}
 R_i x_i=-x_i R_i, \qquad R_i p_i=-p_i R_i, \qquad R_i^2=1.
 \end{equation}
 Notice that the Dunkl derivative \eqref{de} introduces a fictitious angular momentum to the standard harmonic oscillator concerning to odd and even eigenfunction $R_i\psi(x_i)=e_i\psi(x_i)$, so that $e_i=\pm1$, beside the parameter $\mu_i$ plays the role of the angular momentum eigenvalues \cite{operator}.
 \par
In Refs. \cite{Hamiltonian,Hamiltonian1, Hamiltonian2}, it was shown that the Hamiltonian \eqref{eq1} is maximally superintegrable in two- and three-dimensions. Moreover, in making use of the Schwinger construction, their constants of motion and symmetry algebras have uniformly obtained which are the same $u(n), n=2,3$ algebra by the reflection involutions called the Schwinger-Dunkl algebra $sd(n)$. In the same way, we have applied the Jordan-Schwinger representations of three-dimensional orthogonal Cayley-Klein groups \cite{cayley} to provide the constants of motion based on a set of creation and annihilation operators available in the $sl_{-1}(2)$ dynamical symmetry. the generated invariant algebra is a deformation of $so(4)$ by reflection and contraction involutions and it will be called the Jordan-Schwinger-Dunkl algebra $jsd_{\kappa_1\kappa_2}(4)$.
 \par
 In addition to these results, some works in this direction have studied the Dunkl-Coulomb problem based on algebra $so(n)$, one of the simplest of them is the review of the Dunkl-Coulomb problem in the plane \cite{coulomb} in terms of Dunkl operators. Elsewhere, the Dunkl-Laplacian operator is related to $\mathbb{Z}^3_2$ reflection group in the realization of $so(1,2)$ algebra and it is discussed $h$-spherical harmonics \cite{coulomb1,coulomb2,coulomb3}. Also, in generalization of the Dunkl oscillator in the plane \eqref{eq1}, are singular ones associated to the  $su(1,1)$ algebra with a special case of Askey-wilson algebra $AW(3)$ by reflection involutions \cite{wilson,wilson1,wilson2}. These models were mentioned as examples, all are described by Hamiltonians involving the Dunkl-Laplacian operator, that their superintegralbility were shown. Indeed, these works are devoted with the generalization of the Dunkl oscillator in the flat plane to the two- and three-sphere. Recently, superintegrable models by Hamiltonians involving reflection operators are defined on the two- and three-sphere connected to  Bannai-Ito algebra \cite{sphere,sphere1,sphere2,sphere3}. 
 \par
  In this paper, we have used a technique in which Jordan-Schwinger representation is combined with the creation and annihilation operators arising from the $sl_{-1}(2)$ algebra to produce constants of motion and symmetrical algebra. Hereafter, the aim purpose of this work is to show an integrable generalization $\mathcal{H}\kappa$ of the Hamiltonian \eqref{eq1} on the three-dimensional spaces with constant curvature $\kappa\in\{\kappa_1,\kappa_2\}$, and this is exactly what comes from combining the Jordan-Schwinger construction and Dunkl oscillator. In making use of the curvature $\kappa$, it is noteworthy that the three-dimensional spaces $\mathbb{S}^3_{[\kappa_1]\kappa_2}$
  are simultaneously a quadratic form of the metric with signature $\kappa_2$, that is, when  $\kappa_2>0$, it
  recovers a nonrelativistic particle of  whose motion is constrained on the three classical Riemannian
  spaces, while for $\kappa_2<0$, it can be imaged as a relativistic particle constrained on spacetime models.
  In proving that, we will proceed the factorization approach \cite{factor,factor1,factor2}. This
  technique is well known in quantum mechanics, and it is defined in a straightforward way ladder functions
  and shift functions, whose counterpart operators are familiar in quantum mechanics. Hence, we can find
  independent constants of motion, which these constants will characterize the additional symmetries, so that
  our version 3D system is shown to be superintegrable.
  \par
 The paper is organized as follows. In section \ref{sec1}, the generalization of the Euclidean model \eqref{eq1} to curved ones is developed. In such a manner, it will be shown which the $sl_{-1}(2)$ realization and Jordan-Schwinger-Dunkl construction are used to cover symmetries of the total Hamiltonian \eqref{eq1}. It will be seen that the constants of motion are identified as a central extension of the Jordan-Schwinger-Dunkl algebra. Section \ref{sec2} is developed to the study of the curved superintegrable Dunkl Hamiltonians arising in the $jsd_{\kappa_1\kappa_2}(4)$ realization. In section \ref{sec3}, it will be considered the factorization approach from the Dunkl oscillator in geodesic  polar coordinates to examine superintegrability from this point of view. By proceeding with the separation process on the three-dimensional Dunkl oscillator in section \ref{sec4}, the solutions of the energy spectrum and wave functions are obtained. The concluding remarks are given in section \ref{sec5}.
\section{Curved Dunkl oscillator model and Geodesic motion}\label{sec1}
In order to generalize the integrable isotropic Dunkl Hamiltonian on the Euclidean space to $\mathbb{S}^3_{[\kappa_1]\kappa_2}$ in terms of the geodesic polar coordinates, these spaces can be embedded in the linear space $\mathbb{R}^4$ with  
ambient or Weierstrass coordinates  $(x_0,x_1,x_2,x_3)$ subjected to the constraint $x_0^2+\kappa_1 x_1^2+\kappa_1\kappa_2 x_2^2+\kappa_1\kappa_2 x^2_3=1$, so that the origin in $\mathbb{S}^3_{[\kappa_1]\kappa_2}$ corresponds to the point $O=(1,0,0,0)\in \mathbb{R}^4$.
The parameterization of the ambient coordinates in the geodesic polar coordinates  $(r,\theta,\varphi)$ reads \cite{change1,change2}
\begin{equation}\label{eq0}
	\begin{split}
& x_0=C_{\kappa_1}(r), \\
& x_1=S_{\kappa_1}(r)C_{\kappa_2}(\theta),\\
& x_2=S_{\kappa_1}(r)S_{\kappa_2}(\theta)\cos\varphi,\\
& x_3=S_{\kappa_1}(r)S_{\kappa_2}(\theta)\sin\varphi,
	\end{split}
\end{equation}
which on Euclidean space $\mathbb{E}^3$ i.e., when $\kappa_1=0$ and $\kappa_2=1$ leads to
\begin{equation}
x_0=0, \qquad x_1=r\cos\theta, \qquad x_2=r\sin\theta\cos\varphi, \qquad x_3=r\sin\theta\sin\varphi.
\end{equation}
 In this approach, the metric on the curved spaces follows from the usual metric in $\mathbb{R}^4$ divided by the curvature $\kappa_1$ in the form \cite{change1,change2}
 \begin{equation}\label{eq2}
 \mathrm{d}s^2=\frac{1}{\kappa_1}\left(\mathrm{d}x_0^2+\kappa_1\mathrm{d}x_1^2+\kappa_1\kappa_2\mathrm{d}x_2^2+\kappa_1\kappa_2\mathrm{d}x_3^2\right),
 \end{equation}
 which this metric in coordinates \eqref{eq0} is written as
\begin{equation}\label{eq3}
\mathrm{d}s^2=\mathrm{d}r^2+\kappa_2S_{\kappa_1}^2(r)\left(\mathrm{d}\theta^2+S_{\kappa_2}^2(\theta)\mathrm{d}\varphi^2\right).
\end{equation}
In this case, using two real contraction parameters, $\kappa_1$ and $\kappa_2$, we are dealing with a family of Cayley-Klein groups in which are introduced the $\kappa$-dependent cosine and sine functions by \cite{change1,change2}
\begin{equation}
C_{\kappa}(x)=
\begin{cases}
\cos \sqrt{\kappa}x, & \kappa\textgreater 0; \\
1, & \kappa=0; \\
\cosh \sqrt{-\kappa}x, & \kappa \textless 0,
\end{cases} \qquad  \quad
S_{\kappa}(x)=
\begin{cases}
\frac{1}{\sqrt{\kappa}}\sin \sqrt{\kappa}x, & \kappa\textgreater 0; \\
x, & \kappa=0; \\
\frac{1}{\sqrt{-\kappa}}\sinh \sqrt{-\kappa}x, & \kappa\textless 0,
\end{cases} 
\end{equation}
The relations involved for these $\kappa$-dependent functions can be found in \cite{change2,am1,am2,proper}
\begin{align}
\begin{split}
& C_{\kappa}^2 (x)+\kappa S_{\kappa}^2(x)=1, \qquad
 \frac{\mathrm{d}}{\mathrm{d}x}C_{\kappa} (x)=-\kappa S_{\kappa} (x), \qquad
\frac{\mathrm{d}}{\mathrm{d}x}S_{\kappa} (x)=C_{\kappa} (x), \\
& \frac{\mathrm{d}}{\mathrm{d}x}{\rm arc}C_{\kappa}(x)=-\frac{1}{\sqrt{\kappa(1-x^2)}},\quad\frac{\mathrm{d}}{\mathrm{d}x}{\rm arc}S_{\kappa}(x)=\frac{1}{\sqrt{1-\kappa x^2}},\quad
	\frac{\mathrm{d}}{\mathrm{d}x}{\rm arc}T_{\kappa}(x)=\frac{1}{1+\kappa x^2},\\
& \frac{\mathrm{d}}{\mathrm{d}x}T_{\kappa}(x)=\frac{1}{C_{\kappa}^2(x)},\qquad\qquad\quad\frac{\mathrm{d}}{\mathrm{d}x}{\tan^{-1}}(x)=\frac{1}{1+x^2}
\end{split}
\end{align}
and definition of the $\kappa$-tangent is as $T_{\kappa} (x)=S_{\kappa} (x)/C_{\kappa} (x)$. The curved oscillator potential subjected to the governing spaces represented the following expression 
\begin{equation}\label{eq9}
	\mathcal{U}=\frac{1}{2}m\omega^2\left(\frac{1-x_0^2}{\kappa_1x_0^2}\right)=\frac{1}{2}m\omega^2\left(\frac{x_1^2+\kappa_2x_2^2+\kappa_2x_3^2}{x_0^2}\right)=\frac{1}{2}m\omega^2
	T^2_{\kappa_1}(r),
\end{equation}
where is known as the Higgs oscillator potential on the curved spaces \cite{phiggs}.
With this process, in terms of the curved parameters and the geodesic polar coordinates given in \eqref{eq0}, in order to derive the Dunkl oscillator Hamiltonian in the curved space it is necessary to make use of the following helpful relations
	\begin{equation}
	\varphi=\tan^{-1}\left(\frac{x_3}{x_2}\right), \quad \theta={\rm arc}T_{\kappa_2}\left(\frac{\sqrt{x_{2}^2+x_{3}^2}}{x_{1}}\right), \quad r={\rm arc}T_{\kappa_{1}}\left(\frac{\sqrt{x_{1}^2+\kappa_{2}(x_{2}^2+x_{3}^2)}}{x_{0}} \right),
	\end{equation}
	and
	\begin{align}
	\begin{split}
	& \frac{\partial}{\partial x_{0}}=-S_{\kappa_{1}}(r)\frac{\partial}{\partial r},\\[3mm]
	&\frac{\partial}{\partial x_{1}}=C_{\kappa_{1}}(r)C_{\kappa_{2}}(\theta)\frac{\partial}{\partial r}-\frac{S_{\kappa_{2}}(\theta)}{S_{\kappa_{1}}(r)}\frac{\partial}{\partial \theta},\\[3mm]
	& \frac{\partial}{\partial x_{2}}=\kappa_{2} C_{\kappa_{1}}(r)S_{\kappa_{2}}(\theta)\cos(\varphi)\frac{\partial}{\partial r}+\frac{C_{\kappa_{2}}(\theta)\cos(\varphi)}{S_{\kappa_{1}}(r)}\frac{\partial}{\partial \theta}-\frac{\sin(\varphi)}{S_{\kappa_{1}}(r)S_{\kappa_{2}}(\theta)}\frac{\partial}{\partial \varphi},\\[3mm]
	& \frac{\partial}{\partial x_{3}}=\kappa_{2} C_{\kappa_{1}}(r)S_{\kappa_{2}}(\theta)\sin(\varphi)\frac{\partial}{\partial r}+\frac{C_{\kappa_{2}}(\theta)\sin(\varphi)}{S_{\kappa_{1}}(r)}\frac{\partial}{\partial \theta}+\frac{\cos(\varphi)}{S_{\kappa_{1}}(r)S_{\kappa_{2}}(\theta)}\frac{\partial}{\partial \varphi}.
	\end{split}
	\end{align}
	After tendinous calculation, the Dunkl oscillator Hamiltonian can be found as
\begin{equation}\label{eq10}
\mathcal{H}_\kappa=A_r^{\kappa_1}+\frac{B_\theta^{\kappa_2}}{2\kappa_2S_{\kappa_1}^2(r)}+\frac{ C_\varphi}{2\kappa_2S_{\kappa_1}^2(r)S_{\kappa_2}^2(\theta)},
\end{equation}
where
\begin{align}
A_r^{\kappa_1}= &-\frac{1}{2}\frac{\partial^2}{\partial r^2}+\left(\kappa_1 \mu_0 T_{\kappa_1} (r)-\left(1+\mu_1+\mu_2+\mu_3\right)\frac{1}{T_{\kappa_1}(r)}\right)\frac{\partial}{\partial r}+\frac{\kappa_1[\mu_0(1-R_0)+\omega^2]}{2C_{\kappa_1}^2(r)}-\frac{\kappa_1\omega^2}{2} ,\\
B_\theta^{\kappa_2}= & -\frac{\partial^2}{\partial \theta^2}+2\left(\kappa_2\mu_1T_{\kappa_2}(\theta)-\left(\frac{1}{2}+\mu_2+\mu_3\right)\frac{1}{T_{\kappa_2}(\theta)}\right)\frac{\partial}{\partial \theta}+\frac{\kappa_2\mu_1(1-R_1)}{C_{\kappa_2}^2(\theta)},\label{eq17}\\
C_\varphi= & -\frac{\partial^2}{\partial \varphi^2}+2\left(\mu_2\tan(\varphi)-\frac{\mu_3}{\tan(\varphi)}\right)\frac{\partial}{\partial \varphi}+\frac{\mu_2(1-R_2)}{\cos^2(\varphi)}+\frac{\mu_3(1-R_3)}{\sin^2(\varphi)},\label{eq19}
\end{align}
which includes all three-dimensional Riemannian and relativistic spaces, and the reflection operators in geodesic polar variables have the action
\begin{align}
&R_0\psi(r,\theta,\varphi)=\psi(\frac{\pi}{\sqrt{\kappa_1}}-r,\theta,\varphi), \quad R_1\psi(r,\theta,\varphi)=\psi(r,\frac{\pi}{\sqrt{\kappa_2}}-\theta,\varphi), \quad R_2\psi(r,\theta,\varphi)=\psi(r,\theta,\pi-\varphi), \\
& R_3\psi(r,\theta,\varphi)=\psi(r,\theta,-\varphi).\nonumber
\end{align}
Since the reflection operators $R_i(i=0,1,2,3)$ can be viewed as rotations in $O(4)$, The Laplace operator or general Hamiltonian \eqref{eq10} has an expression to functions defined on the unit sphere to show its superintegrability.
The three-dimensional Dunkl oscillator Hamiltonian \eqref{eq1} of the model on the Euclidean space represents a realization of the $sl_{-1}(2)$ algebra through one of the its one-dimensional combinations $H_i$. Indeed, the $sl_{-1}(2)$ algebra, introduced in Refs. \cite{Hamiltonian, sl}, gives the creation/annihilation operators
\begin{equation}\label{eq4}
q_{(i)}^{\pm}=\frac{1}{\sqrt{2}}\left(x_i\pm D^{\mu_i}_i\right),
\end{equation}
and consider $q^0_{(i)}=H_i$ where $i=0,1,2,3$. It is easy to see that these generators together with the reflective operators satisfy the relations defined by algebra $sl_{-1}(2)$ which are in the following form 
\begin{equation}
[q^0_{(i)},q^\pm_{(i)}]=\pm q^\pm_{(i)}, \qquad [q^0_{(i)},R_i]=0, \qquad \{q^+_{(i)},q^-_{(i)}\}=2q^0_{(i)}, \qquad \{q^\pm_{(i)},R_i\}=0,
\end{equation}
the $sl_{-1}(2)$ Casimir oprator $C_{(i)}$ becomes \cite{sl}
\begin{equation}
C_{(i)}=q^+_{(i)}q^-_{(i)}R_i-q^0_{(i)}R_i+R_i/2=-\mu_i.
\end{equation}
For the one-dimensional Dunkl oscillator $q^0_{(i)}=H_i$ of view point the representation theory of $sl_{-1}(2)$, one finds
\begin{equation}
E_i=n_i+\mu_i+1/2,
\end{equation}
provided that $\mu_i>-1/2$. Therefore the complete spectrum of the Dunkl oscillator Hamiltonian \eqref{eq1} is
\begin{equation}
E_N=N+\mu_1+\mu_2+\mu_3+3/2, \qquad N=n_1+n_2+n_3=0,1,\dots
\end{equation}
Now, by glance looking at a subset of real Lie algebras containing in the family of the Cayley-Klein orthogonal algebras \cite{cayley} can be express that operators
\begin{equation}\label{eq5}
\hat{J}_{\mu\nu}(\kappa)=\psi\hat{a}^+J_{\mu\nu}(\kappa)\psi\hat{a}^-,
\end{equation}
provide the Jordan-Schwinger representation of the algebra $so_\kappa(4)$ when $\mu,\nu=0,1,2,3$ and $\mu<\nu$. Hence, under the transformation of creation and annihilation operators \eqref{eq4}, we have
\begin{equation}
\psi\hat{a}^-=\left(q_{(0)}^-,q_{(k)}^-\prod^k_{m=1}\kappa_m^{1/2}\right), \qquad \psi\hat{a}^+=\left(q_{(0)}^+,q_{(k)}^+\prod^k_{m=1}\kappa_m^{-1/2}\right), \qquad k=1,2,3
\end{equation}
which $\psi$ is said to be identical, so that $\psi\hat{a}^-=\hat{a}^-$ and $\psi\hat{a}^+=\hat{a}^+$, whenever $\kappa_m=1$.
It can clearly see that the map of the  algebra $so_\kappa(4)$ is induced in creation and annihilation operators \eqref{eq4}, which has a definition of
\begin{equation}
\psi x_0=x_0, \qquad \psi x_k=x_k \prod_{m=1}^k \kappa_m^{1/2},
\end{equation}
such that $\kappa_3=1$ is considered here.
Thus, the Jordan-Schwinger construction \eqref{eq5} of Lie generators $so_{\kappa_1\kappa_2}(4)$  is shown in making use of the creation and annihilation operators \eqref{eq4}, as follows $(j=2,3)$
\begin{align}
\begin{split}
& \hat{J}_{01}=q_{(0)}^+q_{(1)}^--\kappa_1q_{(0)}^-q_{(1)}^+, \qquad \hat{J}_{1j}=q_{(1)}^+q_{(j)}^--\kappa_2q_{(1)}^-q_{(j)}^+,\\& \hat{J}_{0j}=q_{(0)}^+q_{(j)}^--\kappa_1\kappa_2q_{(0)}^-q_{(j)}^+, \qquad \hat{J}_{23}=q_{(2)}^+q_{(3)}^--q_{(2)}^-q_{(3)}^+.
\end{split}
\end{align}
On the other hand, by sitting the creation and annihilation operators in generators, we obtain each of them in terms of Dunkl derivatives as
\begin{align}\label{eq7}
\begin{split}
& \hat{J}_{01}=x_0 D_1^{\mu_1}-\kappa_1x_1 D_0^{\mu_0}, \qquad \hat{J}_{1j}=x_1 D_j^{\mu_j}-\kappa_2 x_j D_1^{\mu_1},\\
& \hat{J}_{0j}=x_0 D_j^{\mu_j}-\kappa_1\kappa_2x_j D_0^{\mu_0}, \qquad \hat{J}_{23}=x_2 D_3^{\mu_3}-x_3 D_2^{\mu_2},
\end{split}
\end{align}
which can basically be interpreted as Dunkl rotation generators in the ambient coordinates.
Notice that $SO_{\kappa_1,\kappa_2}(4)$ is an isometric group of $diag(+1,\kappa_1,\kappa_1\kappa_2,\kappa_1\kappa_2)$ acting on the linear ambient space $\mathbb{R}^4=(x_0,x_1,x_2,x_3)$ via matrix multiplication. Upon the Cartan decomposition of the algebra $so_{\kappa_1,\kappa_2}(4)=\mathfrak{h}+\mathfrak{p}$ defined by
\begin{equation}
\mathfrak{h}=\langle\hat{J}_{12},\hat{J}_{13},\hat{J}_{23}\rangle=so_{\kappa_2}(3), \qquad \mathfrak{p}=\langle\hat{J}_{01},\hat{J}_{02},\hat{J}_{03}\rangle,
\end{equation}
One can realize $\mathfrak{h}$ as Lie subalgebra which gives homogeneous symmetric spaces in three dimensions, namely $\mathbb{S}^3_{[\kappa_1]\kappa_2}=SO_{\kappa_1,\kappa_2}(4)/SO_{\kappa_2}(3)$, which has constant curvature equal to $\kappa_1$, and $\kappa_2$ plays a role of the signature governing on the  metric  $diag(+1,\kappa_2,\kappa_2)$. So that by being positive $\kappa_2$, it recovers three Riemannian spaces and when it contains the negative value given in a Lorentzian metric. Generally, the metric on $\mathbb{S}^3_{[\kappa_1]\kappa_2}$ is the same as stated in Eq. \eqref{eq2}, showing a map from Euclidean space to Cayley-Klein space. Since the metric \eqref{eq2} is understood in kinetic energy as the ambient velocities that cause the momenta to follow it and Dunkl derivatives have similar behavior via replacing
\begin{equation}\label{eq6}
\partial_i^{\mu_i}\rightarrow \left(\partial_i^{\mu_i}\right)'=g_{ij}\partial_j^{\mu_j}.
\end{equation}
Next, in the calculation of the Dunkl derivatives \eqref{eq6} in the parametrization \eqref{eq0}, the generators \eqref{eq7} in geodesic polar coordinates and their derivatives turn out to be
\begin{align}\label{eq16}
& \hat{J}_{01}=C_{\kappa_2}(\theta)\frac{\partial}{\partial r}-\frac{S_{\kappa_2}(\theta)}{T_{\kappa_1}(r)}\frac{\partial}{\partial \theta}+\frac{\mu_1(1-R_1)}{T_{\kappa_1}(r)C_{\kappa_2}(\theta)}-\kappa_1T_{\kappa_1}(r)C_{\kappa_2}(\theta)\mu_0(1-R_0),\notag\\
& \hat{J}_{02}=\kappa_2S_{\kappa_2}(\theta)\cos\varphi\frac{\partial}{\partial r}+\frac{C_{\kappa_2}(\theta)\cos\varphi}{T_{\kappa_1}(r)}\frac{\partial}{\partial \theta}-\frac{\sin\varphi}{T_{\kappa_1}(r)S_{\kappa_2}(\theta)}\frac{\partial}{\partial \varphi}+\frac{\mu_2(1-R_2)}{T_{\kappa_1}(r)S_{\kappa_2}(\theta)\cos\varphi}-\kappa_1\kappa_2T_{\kappa_1}(r)S_{\kappa_2}(\theta)\cos\varphi\mu_0(1-R_0),\notag\\
& \hat{J}_{03}=\kappa_2S_{\kappa_2}(\theta)\sin\varphi\frac{\partial}{\partial r}+\frac{C_{\kappa_2}(\theta)\sin\varphi}{T_{\kappa_1}(r)}\frac{\partial}{\partial \theta}+\frac{\cos\varphi}{T_{\kappa_1}(r)S_{\kappa_2}(\theta)}\frac{\partial}{\partial \varphi}+\frac{\mu_3(1-R_3)}{T_{\kappa_1}(r)S_{\kappa_2}(\theta)\sin\varphi}-\kappa_1\kappa_2T_{\kappa_1}(r)S_{\kappa_2}(\theta)\sin\varphi\mu_0(1-R_0),\notag\\
& \hat{J}_{12}=\cos\varphi\frac{\partial}{\partial \theta}-\frac{\sin\varphi}{T_{\kappa_2}(\theta)}\frac{\partial}{\partial \varphi}+\frac{\mu_2(1-R_2)}{T_{\kappa_2}(\theta)\cos\varphi}-\kappa_2T_{\kappa_2}(\theta)\cos\varphi\mu_1(1-R_1),\notag\\
& \hat{J}_{13}=\sin\varphi\frac{\partial}{\partial \theta}+\frac{\cos\varphi}{T_{\kappa_2}(\theta)}\frac{\partial}{\partial \varphi}+\frac{\mu_3(1-R_3)}{T_{\kappa_2}(\theta)\sin\varphi}-\kappa_2T_{\kappa_2}(\theta)\sin\varphi\mu_1(1-R_1),\notag\\
& \hat{J}_{23}=\frac{\partial}{\partial \varphi}+\frac{\mu_3(1-R_3)}{\tan\varphi}-\tan\varphi\mu_2(1-R_2).
\end{align}
Direct computations among the the generators of the algebra $so_{\kappa_1,\kappa_2}(4)$ satisfy the commutation relations $(j,k=2,3)$ 
\begin{align}\label{eq8}
&[\hat{J}_{01},\hat{J}_{0j}]=\kappa_1\hat{J}_{1j}(1+2\mu_0R_0), \quad \qquad\qquad\qquad \quad [\hat{J}_{01},\hat{J}_{1j}]=-\hat{J}_{0j}(1+2\mu_1R_1),\\
 &
[\hat{J}_{01},\hat{J}_{23}]=0, \qquad \qquad \qquad\qquad  \  \qquad\qquad\qquad [\hat{J}_{0j},\hat{J}_{0k}]=\kappa_1\kappa_2\epsilon_{0jk}\hat{J}_{23}(1+2\mu_0R_0),\notag\\
&[\hat{J}_{0j},\hat{J}_{1k}]=\kappa_2\delta_{jk}\hat{J}_{01}(1+2\mu_jR_j), \quad \qquad \qquad \qquad [\hat{J}_{0j},\hat{J}_{23}]=-\hat{J}_{0k}(1+2\mu_jR_j),\ j\neq k\notag \\  
&[\hat{J}_{1j},\hat{J}_{23}]=\epsilon_{j1k}\hat{J}_{1k}(1+2\mu_jR_j),\qquad \qquad \qquad \quad \ \  
[\hat{J}_{1j},\hat{J}_{1k}]=\kappa_2\epsilon_{1jk}\hat{J}_{23}(1+2\mu_1R_1),\notag
\end{align}
in which $\delta$ and $\epsilon$ introduce delta- and epsilon-function.
This algebra admits the Casimir  invariant
\begin{align}\label{eq23}
\mathcal{C}_1&=\kappa_2J_{01}^2(1-4\mu_0\mu_1R_0R_1)+\sum_{j=2}^{3}J_{0j}^2(1-4\mu_0\mu_jR_0R_j)+\kappa_1\sum_{j=2}^{3}J_{1j}^2(1-4\mu_1\mu_jR_1R_j)+\kappa_1\kappa_2J_{23}^2(1-4\mu_2\mu_3R_2R_3),\notag\\
\mathcal{C}_2&=\kappa_2J_{01}^2+\sum_{j=2}^{3}J_{0j}^2+\kappa_1\sum_{j=2}^{3}J_{1j}^2+\kappa_1\kappa_2J_{23}^2+2\kappa_1\kappa_2\sum_{i=0}^{3}\mu_i(1-R_i)+2\kappa_1\sum_{0\leq i<j\leq3}\mu_i\mu_j(1-R_iR_j),
\end{align}
where $\mathcal{C}_1$ is associated to the Killing-Cartan form.
On the other hand, the commutation and anticommutation relations of the generators of this algebra with the reflections operators provide
\begin{equation}\label{eq88}
[\hat{J}_i,R_i]=0, \qquad \{\hat{J}_j,R_k\}=0.
\end{equation}
In this way, we can read the invariant algebra resulting from the commutation relations given in Eqs. \eqref{eq8} and \eqref{eq88} as Jordan-Schwinger-Dunkl algebra $jsd_{\kappa_1\kappa_2}(4)$, which is a deformation of the $so(4)$ usual Lie algebra by involutions.
\section{Superintegrability}\label{sec2}
This section induces from the previous section that the three-dimensional Dunkl oscillator model is maximally superintegrable.
Indeed, the free Hamiltonian $\mathcal{H}_\kappa$ with $\omega=0$, which is given in \eqref{eq10} arising the Jordan-Schwinger-Dunkl procedure exhibits three integrals of the motion associated to the rotation generators with $j=2,3$
\begin{align} L_{1j}=\hat{J}^2_{1j}, \qquad \qquad \qquad \qquad \ L_{23}=\hat{J}^2_{23}
\end{align}
where $\hat{J}_{ij}$ are the functions \eqref{eq16}. 
The constants of the motion $L_{1j}$ and $L_{23}$ do not commute with each other. To find quantities in involution, we define another set from integrals above as follows
\begin{align}
L_{123}&=L_{12}+L_{13}+\kappa_2L_{23}+\kappa_2\sum_{i=1}^{3}\mu_i(1-R_i)+2\sum_{1\leq i<j\leq3}\mu_i\mu_j(1-R_iR_j)\\&=\frac{\partial^2}{\partial\theta^2}-2\left[\kappa_2\mu_1T_{\kappa_2}^2(\theta)-\left(\frac{1}{2}+\mu_2+\mu_3\right)\frac{1}{T_{\kappa_2}^2(\theta)}\right]\frac{\partial}{\partial\theta}-\frac{\kappa_2\mu_1(1-R_1)}{C^2_{\kappa_2}(\theta)}-\frac{C_\varphi}{S^2_{\kappa_2}(\theta)}
\end{align}
in other words, we have
\begin{equation} L_{123}=-\Delta_{\mathbb{S}^2_{\kappa_2}}+\kappa_2\sum_{i=1}^{3}\mu_i(1-R_i)+2\sum_{1\leq i<j\leq3}\mu_i\mu_j(1-R_iR_j)
\end{equation}
and we define
\begin{equation}
 \Delta_{\mathbb{S}^2_{\kappa_2}}= B_\theta^{\kappa_2}+\frac{C_\varphi}{S^2_{\kappa_2}(\theta)}\label{eq43}
\end{equation}
in which $B_\theta^{\kappa_2}$ and $C_\varphi$ are expressed in \eqref{eq17} and \eqref{eq19}, respectively.
Moreover, the operator $\Delta_{\mathbb{S}^2_{\kappa_2}}$ is called the Dunkl-Laplacian operator and is related to the Casimir of the rotation subalgebra $\mathfrak{h}=jsd_{\kappa_2}(3)$.
It is directly checked that the operators
 $(L_{12}, L_{123},\mathcal{H}_\kappa)$ are the conserved quantities of the curved Dunkl Hamiltonian in involution together. The same
is for the set $(L_{23}, L_{123},\mathcal{H}_\kappa)$.
In generally, the four constants of motion $(L_{12}, L_{23}, L_{123},\mathcal{H}_\kappa)$ are algebrically dependent. Hence it follows that the three-dimensional Dunkl Hamiltonian $\mathbb{S}^3_{\kappa_1\kappa_2}$ is maximally superintegrable.
\par
In the other hand, the $\hat{J}_{0i}$'s functions \eqref{eq16}, which represent the generators of translation construct other constants of motion as
\begin{equation}\label{eq24}
L_{0j}=\hat{J}^2_{0j}, \qquad L_{01}=\hat{J}^2_{01}
\end{equation}
Therefore, a combination of transmission and rotation generators provides
\begin{align}\label{eq22}
\begin{split}
\kappa_2 L_{01}&+L_{02}+L_{03}+\kappa_1L_{12}+\kappa_1L_{13}+\kappa_1\kappa_2L_{23}+2\kappa_1\kappa_2\sum_{i=0}^{3}\mu_i(1-R_i)+2\kappa_1\sum_{0\leq i<j\leq3}\mu_i\mu_j(1-R_iR_j)\\
& =\kappa_2L_{01}+L_{02}+L_{03}+\kappa_1L'_{123}
\end{split}
\end{align}
It is observed that combination \eqref{eq22} creates a constraint as follows
\begin{equation}
-2\kappa_2\mathcal{H}_\kappa=\kappa_2L_{01}+L_{02}+L_{03}+\kappa_1L'_{123}.
\end{equation}
Evidently, the integrals of the motion $(L_{01}, L_{02}, L_{03}, L_{12}, L_{23}, L_{123},\mathcal{H}_\kappa)$ behavior as algebraically dependent in mutual involution. Finally, each function $L_{0i} (i =1,2,3)$ given in \eqref{eq24} is in involution with $\mathcal{H}_\kappa$,   which is said to be another set of for the proof of the maximal superintegrability of $\mathcal{H}_\kappa$.
In this respect, notice that, under \eqref{eq16}, the kinetic energy is related to the Casimir \eqref{eq23} by
\begin{equation}
\mathcal{C}=-\left(\mathcal{T}-4\sum_{0\leq i<j\leq3}\mathcal{T}_{ij}\mu_i\mu_jR_iR_j\right),
\end{equation}
where $\mathcal{T}$ read as the kinetic energy of a free particle in the absence of Dunkl space, and it is similar to what is given by the classical state which has been used.
However, since the three-dimensional Dunkl oscillator Hamiltonain \eqref{eq1} contains first order terms in the derivative, let us consider the following gauge transformation introduced in \cite{gauge} for each dimension separately 
\begin{equation}
\tilde{H}_i=|x_i|^{\mu_i}H_i|x_i|^{-\mu_i}.
\end{equation}
Under this transformation, the Dunkl-Laplacian operator available in Eq. \eqref{eq1} takes the simple form 
\begin{equation}
\Delta_D=\tilde{D}_x^{\mu_x}+\tilde{D}_y^{\mu_y}+\tilde{D}_z^{\mu_z},
\end{equation}
which is reduced to the usual Laplacian $\Delta=\partial^2_x+\partial^2_y+\partial^2_z$ and the centrifugal terms with reflections of $R_i$. Thus, the total deformed Hamiltonian from the one-dimensional Hamiltonians is converted to the following form $(i=x,y,z)$
\begin{equation}\label{eq25}
\tilde{H}=\frac{1}{2}\left(-\Delta+\sum_{i=1}^{3}\frac{\mu_i(\mu_i-R_i)}{x_i^2}+\omega^2 x_i^2\right),
\end{equation}
and the Hamiltonian \eqref{eq25} reduces to the harmonic oscillator model when $\mu_i=0$. Now, in using the geodesic polar coordinates \eqref{eq0} and with a transition from $\tilde{\mathcal{H}}^{SW}_\kappa=\tilde{\mathcal{H}}_\kappa-\frac{\kappa_1\mu_0(\mu_0-R_0)}{2}$, we have for the Hamiltonain \eqref{eq25} 
\begin{equation}\label{eq26}
\tilde{\mathcal{H}}^{SW}_\kappa=\frac{1}{2}\left[-\Delta_{LB}+\omega'^2T^2_{\kappa_1}(r)+\frac{1}{S^2_{\kappa_1}(r)}\left(\frac{\mu_1(\mu_1-R_1)}{C^2_{\kappa_2}(\theta)}+\frac{\mu_2(\mu_2-R_2)}{\kappa_2S^2_{\kappa_2}(\theta)\cos^2\varphi}+\frac{\mu_3(\mu_3-R_3)}{\kappa_2S^2_{\kappa_2}(\theta)\sin^2\varphi}\right)\right],
\end{equation}
where $\omega'^2=\omega^2+\mu_0(\mu_0-R_0)$ and $\Delta_{LB}$ is the Laplace-Beltrami operator on curved spaces $\mathbb{S}^3_{[\kappa_1]\kappa_2}$ with curvature parameters $\kappa_1,\kappa_2$ \cite{LB1,LB2}.
We obtain the Laplace-Beltrami is parametrized by the local coordinates $(r, \theta, \varphi)$ on the curved space in the following form
\begin{equation}
\Delta_{LB}=-\frac{\hbar^2}{2m}\left[\frac{1}{S^2_{\kappa_1}(r)}\frac{\partial}{\partial r}S^2_{\kappa_1}(r)\frac{\partial}{\partial r}+\frac{1}{\kappa_2 S_{\kappa_1}^2(r)}\left(\frac{1}{S_{\kappa_2}(\theta)}\frac{\partial}{\partial \theta}S_{\kappa_2}(\theta)\frac{\partial}{\partial \theta}+\frac{1}{ S_{\kappa_2}^2(\theta)}\frac{\partial^2}{\partial \varphi^2}\right)\right].
\end{equation}
Obviously, in Riemannian space ($\kappa_2>0$), when $\kappa_1>0$ this Hamiltonian recovers a system defined on $\mathbb{S}^3$, and when $\kappa_1<0$, it can be expressed on $\mathbb{H}^3$. On the other hand, Hamiltonian \eqref{eq10} for $\kappa_1=0$ leads to the Hamiltonian  of the Euclidean isotropic oscillator $\mathbb{E}^3$. While $\kappa_2<0$, it covers the three spaces involved in the Lorentz metric. We do not consider $\kappa_2=0$, because our purpose here is to study the superintegral systems. 
\par
Now if we select the superintegrable potential on the space $\mathbb{S}^3_{[\kappa_1]\kappa_2}$ as an expression of the superposition of a generalized k-dependent Dunkl oscillator with angular frequency $\omega'$, and the centrifugal barriers $\mu_i(1-R_i)$. So, there is our introduced potential in the form
\begin{align}\label{eq31}
\mathcal{U}^{SW}=\omega'^2T^2_{\kappa_1}(r)+\frac{1}{S^2_{\kappa_1}(r)}\left(\frac{\mu_1(\mu_1-R_1)}{C^2_{\kappa_2}(\theta)}+\frac{\mu_2(\mu_2-R_2)}{\kappa_2S^2_{\kappa_2}(\theta)\cos^2\varphi}+\frac{\mu_3(\mu_3-R_3)}{\kappa_2S^2_{\kappa_2}(\theta)\sin^2\varphi}\right),
\end{align}
which is known as Smorodinsky-Winternitz (SW) potential \cite{sw,sw1,sw2,sw3,sw4}. In Refs. \cite{aniso1,aniso2,aniso3,higgs,higgs1}, the Higgs oscillator is represented as a curved analogous of the Euclidean oscillator, which is a curved integrable generalization with centrifugal terms added to the general Hamiltonian.
Thus, this provides a demonstration of the Higgs-Dunkl curved oscillator Hamiltonian on six spaces with Rosochatius
terms $\mu_i(\mu_i-R_i)$. In fact, here we need to say that Hamiltonian \eqref{eq26} displays the generalization of the three-dimensional SW system, $\tilde{\mathcal{H}}^{SW}_\kappa=\mathcal{T}+\mathcal{U}^{SW}$, to the spaces $\mathbb{S}^3_{[\kappa_1]\kappa_2}$. The SW systems are known as the maximally superintegrable system on the low-dimensional spaces.
The operators $\hat{J}_{\mu\nu}$ exiting in \eqref{eq16} are reduced to the familiar angular momentum generators
\begin{align}\label{eq35}
& \hat{J}_{01}'=C_{\kappa_2}(\theta)\frac{\partial}{\partial r}-\frac{S_{\kappa_2}(\theta)}{T_{\kappa_1}(r)}\frac{\partial}{\partial \theta}\notag\\
& \hat{J}_{02}'=\kappa_2S_{\kappa_2}(\theta)\cos\varphi\frac{\partial}{\partial r}+\frac{C_{\kappa_2}(\theta)\cos\varphi}{T_{\kappa_1}(r)}\frac{\partial}{\partial \theta}-\frac{\sin\varphi}{T_{\kappa_1}(r)S_{\kappa_2}(\theta)}\frac{\partial}{\partial \varphi}\notag\\
& \hat{J}_{03}'=\kappa_2S_{\kappa_2}(\theta)\sin\varphi\frac{\partial}{\partial r}+\frac{C_{\kappa_2}(\theta)\sin\varphi}{T_{\kappa_1}(r)}\frac{\partial}{\partial \theta}+\frac{\cos\varphi}{T_{\kappa_1}(r)S_{\kappa_2}(\theta)}\frac{\partial}{\partial \varphi}\notag\\
& \hat{J}_{12}'=\cos\varphi\frac{\partial}{\partial \theta}-\frac{\sin\varphi}{T_{\kappa_2}(\theta)}\frac{\partial}{\partial \varphi}\notag\\
& \hat{J}_{13}'=\sin\varphi\frac{\partial}{\partial \theta}+\frac{\cos\varphi}{T_{\kappa_2}(\theta)}\frac{\partial}{\partial \varphi}\notag\\
& \hat{J}_{23}'=\frac{\partial}{\partial \varphi}
\end{align}
which covers the algebra $so_{\kappa_1,\kappa_2}(4)$ commutation relations $(j,k=2,3)$ 
\begin{align}
\begin{split}
&[\hat{J}'_{01},\hat{J}'_{0j}]=\kappa_1\hat{J}'_{1j}, \quad \qquad  [\hat{J}'_{01},\hat{J}'_{1j}]=-\hat{J}'_{0j}, \qquad \qquad
[\hat{J}'_{01},\hat{J}'_{23}]=0, \qquad      \quad \ \ [\hat{J}'_{0j},\hat{J}'_{0k}]=\kappa_1\kappa_2\epsilon_{0jk}\hat{J}'_{23},\\
&[\hat{J}_{0j},\hat{J}_{1k}]=\kappa_2\delta_{jk}\hat{J}_{01},\qquad [\hat{J}'_{0j},\hat{J}'_{23}]=-\hat{J}'_{0k},\ j\neq k \quad \ \ [\hat{J}'_{1j},\hat{J}'_{23}]=\epsilon_{j1k}\hat{J}'_{1k},  \quad \ \  
[\hat{J}'_{1j},\hat{J}'_{1k}]=\kappa_2\epsilon_{1jk}\hat{J}'_{23},
\end{split}
\end{align}
and the Casimir operator $\mathcal{C}'$ is given by
\begin{equation}\label{eq36}
\mathcal{C}'=\kappa_2J_{01}'^2+\sum_{j=2}^{3}J_{0j}'^2+\kappa_1\sum_{j=2}^{3}J_{1j}'^2+\kappa_1\kappa_2J_{23}'^2.
\end{equation}
 The total SW Hamiltonian \eqref{eq26} with the potential \eqref{eq31} has resulted in the translation and Lorentz-rotation generators, respectively 
\begin{equation}
\tilde{L}_{01}=-J_{01}'^2+\frac{x_1^2}{x_0^2}\omega'^2+\frac{x_0^2}{x_1^2}\mu_1(\mu_1-R_1), \quad \tilde{L}_{0j}=-J_{0j}'^2+\kappa_2^2\frac{x_j^2}{x_0^2}\omega'^2+\frac{x_0^2}{x_j^2}\mu_j(\mu_j-R_j),
\end{equation}
and 
\begin{equation}
\tilde{L}_{1j}=-J_{1j}'^2+\kappa_2\frac{x_j^2}{x_1^2}\mu_j(\mu_j-R_j)+\frac{x_1^2}{x_j^2}\mu_1(\mu_1-R_1), \quad \tilde{L}_{23}=-J_{23}'^2+\frac{x_3^2}{x_2^2}\mu_2(\mu_2-R_2)+\frac{x_2^2}{x_3^2}\mu_3(\mu_3-R_3),
\end{equation}
or in another representation from the geodesic polar  space \eqref{eq0} are explicitly read
\begin{align}
\begin{split}
&\tilde{L}_{01}=-J_{01}'^2+\frac{\mu_1(\mu_1-R_1)}{T_{\kappa_1}^2(r)C_{\kappa_2}^2(\theta)}+T_{\kappa_1}^2(r)C_{\kappa_2}^2(\theta)\omega'^2,\\
&\tilde{L}_{13}=-J_{02}'^2+\frac{\mu_2(\mu_2-R_2)}{T_{\kappa_1}^2(r)S_{\kappa_2}^2(\theta)\cos^2\varphi}+\kappa_2^2T_{\kappa_1}^2(r)S_{\kappa_2}^2(\theta)\cos^2\varphi\omega'^2,\\
& \tilde{L}_{23}=-J_{03}'^2+\frac{\mu_3(\mu_3-R_3)}{T_{\kappa_1}^2(r)S_{\kappa_2}^2(\theta)\sin^2\varphi}+\kappa_2^2T_{\kappa_1}^2(r)S_{\kappa_2}^2(\theta)\sin^2\varphi\omega'^2,\\
&\tilde{L}_{12}=-J_{12}'^2+\kappa_2\frac{\mu_2(\mu_2-R_2)}{T_{\kappa_2}^2(\theta)\cos^2\varphi}+T_{\kappa_2}^2(\theta)\cos^2\varphi\mu_1(\mu_1-R_1),\\
&\tilde{L}_{13}=-J_{13}'^2+\kappa_2\frac{\mu_3(\mu_3-R_3)}{T_{\kappa_2}^2(\theta)\sin^2\varphi}+T_{\kappa_2}^2(\theta)\sin^2\varphi\mu_1(\mu_1-R_1),\\
& \tilde{L}_{23}=-\partial_\varphi^2+\frac{\mu_3(\mu_3-R_3)}{\tan^2\varphi}+\tan^2\varphi\mu_2(\mu_2-R_2).
\end{split}
\end{align}
These generators do not commute all together. Hence, rotational generators define another quantity as
\begin{align}
\begin{split}
\tilde{L}_{123}&=\tilde{L}_{12}+\tilde{L}_{13}+\kappa_2\tilde{L}_{23}+\kappa_2\sum_{i=1}^{3}\mu_i(\mu_i-R_i)\\
&=-\frac{1}{S_{\kappa_2}(\theta)}\partial_\theta S_{\kappa_2}(\theta)\partial_\theta+\frac{\kappa_2\mu_1(\mu_1-R_1)}{C^2_{\kappa_2}(\theta)}+\frac{1}{S^2_{\kappa_2}(\theta)}\left(-\partial_\varphi^2+\frac{\mu_2(\mu_2-R_2)}{\cos^2\varphi}+\frac{\mu_3(\mu_3-R_3)}{\sin^2\varphi}\right),
\end{split}
\end{align}
which in addition to having integrability property, it is in involution with them  and derive from the Casimir of the rotation subalgebra $so_{\kappa_2}(3)$. Therefore, the direct result of the integrability of the angular momentum shows that rotation generators can find in two separate equations, where each of them depends on only a coordinate, namely
\begin{align}
\begin{split}
& \tilde{L}_{23}=-\partial_\varphi^2+\frac{\mu_3(\mu_3-R_3)}{\tan^2\varphi}+\tan^2\varphi\mu_2(\mu_2-R_2),\\
& \tilde{L}_{123}=-\frac{1}{S_{\kappa_2}(\theta)}\partial_\theta S_{\kappa_2}(\theta)\partial_\theta+\frac{2\kappa_2\mu_1(\mu_1-R_1)}{C^2_{\kappa_2}(\theta)}+\frac{1}{S^2_{\kappa_2}(\theta)}\left(\tilde{L}_{23}+\sum_{j=2}^{3}\mu_j(\mu_j-R_j)\right).
\end{split}
\end{align}
So, one can say that the functions $\{\tilde{L}_{12},\tilde{L}_{123}\}$ and $\{\tilde{L}_{23},\tilde{L}_{123}\}$ are mutually in involution. Now, a combination of transition generators $\tilde{L}_{0i}$ and  extracted integral of the motion, that is $\tilde{L}_{123}$, is constrained to 
\begin{equation}
2\kappa_2\mathcal{H}^{SW}_\kappa=\kappa_2\tilde{L}_{01}+\tilde{L}_{02}+\tilde{L}_{03}+\kappa_1\tilde{L}_{123}.
\end{equation}
In this way, each transition function $\tilde{L}_{0i}$ is commuted with $\mathcal{H}_{\kappa}^{SW}$, and the set of functions $\{\tilde{L}_{0i}, \tilde{L}_{12},\tilde{L}_{23},\tilde{L}_{123},\mathcal{H}_{\kappa}^{SW}\}$ operates independently. So, by direct calcul	ation, seperintegrability of $\mathcal{H}_{\kappa}^{SW}$ is guaranteed and $\mathcal{H}_{\kappa}^{SW}$, is read  as a superintegrable Hamiltonian.
In this respect, notice that, under \eqref{eq35}, the kinetic energy is related to the Casimir $\mathcal{C}'$ \eqref{eq36} of the Jordan-Schwinger algebra by $\mathcal{C}'=-2\kappa_2\mathcal{T}$.
\section{Representation of the Factorization Method on the curved Dunkl oscillator and their exact solutions}\label{sec3}
It is clear that the Hamiltonian $\hat{\mathcal{H}}_\kappa$ in the geodesic polar variables takes the following eigenvalue equation
\begin{equation}\label{eq37}
\hat{\mathcal{H}}_\kappa \Psi_\kappa(r,\theta,\varphi)=\mathcal{E}_\kappa \Psi_\kappa(r,\theta,\varphi),
\end{equation}
which is separable in the mentioned coordinate system. Upon considering  $\Psi_\kappa(r,\theta,\varphi)=R_{\kappa_1}(r)\Theta_{\kappa_2}(\theta)\Phi (\varphi)$, one finds the
system of ordinary equations
\begin{equation}
\hat{\mathcal{H}}_\kappa R_{\kappa_1}(r)=\mathcal{E}_\kappa R_{\kappa_1}(r), \qquad B_{\kappa_2}^\theta \Theta_{\kappa_2}(\theta)=L \Theta_{\kappa_2}(\theta), \qquad C_\varphi \Phi(\varphi)=\ell_z \Phi(\varphi).
\end{equation}
Now to find the factorized solutions \cite{shi}, let us study each of the one-dimensional systems located in \eqref{eq10}. Once again, we go through the same procedure alone for $\hat{\mathcal{H}}_{\kappa_2}^\theta$. In this way, the Hamiltonian $\hat{\mathcal{H}}_\kappa$ is satisfied ladder operators and shift
operators formalism. We also will present that these operators result in additional symmetries for $\hat{\mathcal{H}}_\kappa$. Also, it is shown that Eq. \eqref{eq10} in geodesic polar coordinates be superintegrable based on the factorization formalism and solutions have been presented to it. .
\subsection{first set of the quantum symmetries $\hat{\mathcal{H}}_{\kappa_2}^{\theta\varphi}$}
In this subsection, let us consider a set of factorizable one-dimensional Hamiltonians for $\hat{\mathcal{H}}_{\kappa_2}^{\theta\varphi}$, labeled by parameters of the angular variables $\theta$ and $\varphi$, having the form \eqref{eq43}. Thus, we can write  the factorization approach to obtain the ladder operators $C_{\kappa,\epsilon}^\pm$ for the Hamiltonian $C_\varphi$ given in \eqref{eq19}. Then, in making use of the transformation
\begin{equation}\label{eq15}
	\Phi(\varphi)=\frac{\nu(\varphi)}{\cos(\varphi)^{\mu_2}(\sin(\varphi))^{\mu_3}}, 
\end{equation}
and removing the
first-order derivative for the Hamiltonian \eqref{eq19}, the simple expression remains as
\begin{equation}\label{eq12}
	C_\varphi=-\frac{\partial^2}{\partial \varphi^2}+\frac{\mu_2(\mu_2-R_2)}{\cos^2(\varphi)}+\frac{\mu_3(\mu_3-R_3)}{\sin^2(\varphi)}-(\mu_2+\mu_3)^2.
\end{equation}
Remark that $C_\varphi$ is just the quantum P\"{o}schl-Teller-Dunkl Hamiltonian \cite{poschl} written in $\kappa-$dependent functional version with reflections of $R_2,R_3$. Moreover, it will be helpful to use the notation valid for general eigenvalues in the form
\begin{equation}
	\left(C_\varphi+(\mu_2+\mu_3)^2\right)\nu(\varphi)=\left(\ell_z+(\mu_2+\mu_3)^2\right)\nu(\varphi)=\epsilon^2\nu(\varphi).
\end{equation}
Now to get the ladder operators, we will multiply equation \eqref{eq12} by $\cos^2\varphi$ to obtain
\begin{equation}\label{eq122}
	\left(-\cos^2(\varphi)\frac{\partial^2}{\partial \varphi^2}+\frac{\left(c_2^2-\tfrac{1}{4}\right)}{\tan^2(\varphi)}-\epsilon ^2\cos^2(\varphi)\right)\nu(\varphi)=-\left(c_1^2-\frac{1}{4}\right)\nu(\varphi).
\end{equation}
Hence, we able to introduce the new differential operator in a simpler form of Eq. \eqref{eq122} as
\begin{equation}\label{eq13}
	c_\varphi=-\cos^2(\varphi)\frac{\partial^2}{\partial \varphi^2}+\frac{\left(c_2^2-\tfrac{1}{4}\right)}{\tan^2(\varphi)}-\epsilon ^2\cos^2(\varphi),
\end{equation}
then, we are able to factorize \eqref{eq13} in terms of two first-order differential operators in the usual way
\begin{equation}\label{eq20}
	c_\varphi=C_{\epsilon}^- C_{\epsilon}^++\xi_{\epsilon},
\end{equation}
where
\begin{align}\label{eq14}
\begin{split}
	C_{\epsilon}^-=&\ \cos(\varphi)\frac{\partial}{\partial \varphi}+\frac{\left(c_2+\tfrac{1}{2}\right)}{\sin(\varphi)}+(\epsilon+1)\sin(\varphi), \\
	C_{\epsilon}^+=&-\cos(\varphi)\frac{\partial}{\partial \varphi}+\frac{\left(c_2+\tfrac{1}{2}\right)}{\sin(\varphi)}+\epsilon sin(\varphi),
	\end{split}
\end{align}
and
\begin{equation}
	\xi_{\epsilon}=-\left(c_2+ \epsilon+\frac{3}{2}\right)\left( c_2+ \epsilon+\frac{1}{2}\right).
\end{equation}
Here, the eigenvalues of the reflections $R_2$ and $R_3$with $e_2$ and $e_3$ indicator functions determine the following quantities $c_1$ and $c_2$ by acting on $\nu(\varphi)$, namely
\begin{equation}
	\begin{cases}
		c_1^+=\mu_2-\frac{1}{2}, \qquad \quad \textrm{if} \quad e_2=1\\
		c_1^-=\mu_2+\frac{1}{2}, \quad \qquad  \textrm{if} \quad e_2=-1
	\end{cases}, 
	\begin{cases}
		c_2^+=\mu_3-\frac{1}{2}, \qquad \textrm{if} \quad e_3=1\\
		c_2^-=\mu_3+\frac{1}{2}, \qquad \textrm{if} \quad e_3=-1
	\end{cases},
\end{equation}
The action of the ladder operators \eqref{eq14} on the differential operator of $c_{\epsilon,\varphi}$ is
\begin{equation}
	C_{\epsilon}^+c_{\epsilon,\varphi}=c_{\epsilon+1,\varphi}C_{\epsilon}^+, \qquad C_{\epsilon}^-c_{\epsilon+1,\varphi}=c_{\epsilon,\varphi}C_{\epsilon}^-.
\end{equation}
Therefore, as a result of acting on differential operator of $c_{\epsilon,\varphi}$, their commutator relations are as follows
\begin{equation}
	[c_{\epsilon,\varphi},C_{\epsilon}^-]=-C_{\epsilon}^-\left(2(c_2+\epsilon+1)+1\right),\qquad [c_{\epsilon,\varphi},C_{\epsilon}^+]=\left(2(c_2+\epsilon+1)+1\right)C_{\epsilon}^+,
\end{equation}
where the set of answers from the commutator relations is explicitly a function dependent on one curvature parameter $\kappa$ and the other constants involved in the Dunkl oscillator $(\mu_2,\mu_3)$.
On the other hand, to find the shift operators of $B_{\kappa_2}^\theta$ given in Eq. \eqref{eq17}, it is clearly seen from the expression \eqref{eq17}, the one-dimensional Hamiltonian contains a first-order term of the derivative and it can be eliminated by proposing
\begin{equation}\label{eq32}
\Theta_{\kappa_2}(\theta)=\frac{\chi_{\kappa_2}(\theta)}{(C_{\kappa_2}(\theta))^{\mu_1}(S_{\kappa_2}(\theta))^{(\mu_2+\mu_3+\frac{1}{2})}},
\end{equation}
under this transformation, the Hamiltonian \eqref{eq17} takes the simple expression
\begin{equation}\label{eq27}
\left(-\frac{\partial^2}{\partial\theta^2}+\frac{\kappa_2\mu_1(\mu_1-R_1)}{C_{\kappa_2}^2(\theta)}+\frac{\epsilon^2-\tfrac{1}{4}}{S_{\kappa_2}^2(\theta)}-\kappa_2\left(\mu_1+\mu_2+\mu_3+\frac{1}{2}\right)^2\right)\chi_{\kappa_2}(\theta)=L\chi_{\kappa_2}(\theta).
\end{equation}
Whenever, inspired by
that the Hamiltonian \eqref{eq27} is part of the first P\"{o}schl-Teller Hamiltonians \cite{poschl2} in the following form 
\begin{equation}\label{eq299}
-\frac{\partial^2}{\partial\theta^2}+\frac{\kappa_2d_1(d_1-1)}{C_{\kappa_2}^2(\theta)}+\frac{d_2(d_2-1)}{S_{\kappa_2}^2(\theta)}-\kappa_2\left(\mu_1+\mu_2+\mu_3+\frac{1}{2}\right)^2.
\end{equation}
Then, we are able to factorize \eqref{eq299} in terms of two first-order differential operators in the usual way
\begin{equation}\label{eq21}
	B_\theta^{\kappa_2}=D_{\kappa_2}^+ D_{\kappa_2}^-+\xi_{\kappa_2},
\end{equation}
where yields the shift functions
\begin{equation}\label{eq144}
	D_{\kappa_2}^\pm=\mp\frac{\partial}{\partial \theta}-\kappa_2 d_1T_{\kappa_2}(\theta)+\frac{d_2}{T_{\kappa_2}(\theta)},
\end{equation}
with a spectrum of
\begin{equation}
	\xi_{\kappa_2}=\kappa_2\left[(d_1+d_2)^2+(\mu_1+\mu_2+\mu_3+\tfrac{1}{2})^2\right].
\end{equation}
It is understood that a direct computation on $\mathcal{X}_{\kappa_2}(\theta)$ shows that the reflection operator $R_1=\pm1$ leads to
\begin{equation}
 \begin{cases}
		d_1^+=\mu_1, \quad\ \qquad\ \textrm{if} \quad e_1=1\\
		d_1^-=\mu_1+1, \qquad \textrm{if} \quad e_1=-1
	\end{cases},
\end{equation}
and we can find $d_2=\frac{1}{2}\left[2\epsilon+1\right]$. Accordingly, the shift operators $D_\kappa^\pm$ satisfy the commutation relation
\begin{equation}
	[D_{\kappa_2}^-,D_{\kappa_2}^+]=\frac{\kappa_2 d_1}{C_{\kappa_2}^2(\theta)}+\frac{ d_2}{S_{\kappa_2}^2(\theta)}, \qquad [\mathcal{H}_\kappa,D_{\kappa_2}^\pm]=\pm\left(\frac{\kappa_2 d_1}{C_{\kappa_2}^2(\theta)}+\frac{ d_2}{S_{\kappa_2}^2(\theta)}\right)D_{\kappa_2}^\pm.
\end{equation}
Now, we can construct a pair of the additional symmetry operators in this class of the angular coordinates characterized by
\begin{equation}
[\mathcal{H}_{\kappa_2}^{\theta\varphi},X_\kappa^\pm]=0,
\end{equation}
which includes combinations of operators \eqref{eq14} and \eqref{eq144} , that can be defined as
\begin{equation}\label{eq51}
X_\kappa^\pm=(C_{\epsilon}^\pm)(D_{\kappa_2}^\pm)=\pm\mathcal{O}_\kappa \hat{\epsilon}+\mathcal{W}_\kappa.
\end{equation}
The real polynomial symmetries $\mathcal{W}_\kappa$ and $\mathcal{O}_\kappa$ are defined as orders of the momentum operators. 
Due to the factorization \eqref{eq20} and \eqref{eq21}, the products $X_\kappa^+X_\kappa^-$ and $X_\kappa^-X_\kappa^+$ are contained in the Hamiltonians $B_\theta^{\kappa_2},C_\varphi$, so the set of symmetries $(\mathcal{H}_{\kappa_2}^{\theta\varphi},C_\varphi,X_\kappa^\pm)$ is not algebraically independent.
\subsection{second set of the quantum symmetries $\hat{\mathcal{H}}_\kappa$}
Here, we shall use the factorization approach to obtain the ladder operators $A_{\kappa_2,\varrho}^\pm$ for $B_{\kappa_2}^\theta$.  
In order to find the ladder operators $B_\theta^{\kappa_2}$ for the zenithal Hamiltonian \eqref{eq17} we can use of the calculations of the previous section. 
Moreover, it will be helpful to use the notation valid for general eigenvalues from Eq. \eqref{eq27}  in the form
\begin{equation}
\left(B_{\kappa_2}^\theta+\kappa_2(\mu_1+\mu_2+\mu_3+1/2)^2\right)\chi_{\kappa_2}(\theta)=\left(L+\kappa_2(\mu_1+\mu_2+\mu_3+1/2)^2\right)\chi_{\kappa_2}(\theta)=\kappa_2\hat{\varrho}_{\kappa_2}^2\chi_{\kappa_2}(\theta).
\end{equation}
Now to get the ladder operators, we will multiply Eq. \eqref{eq27} by $C_{\kappa_2}^2(\theta)$ to obtain
\begin{equation}\label{eq28}
\left(-C_{\kappa_2}^2(\theta)\frac{\partial^2}{\partial\theta^2}+\frac{\left(\epsilon^2-\tfrac{1}{4}\right)}{T^2_{\kappa_2}(\theta)}-\kappa_2c_{\kappa_2}^2(\theta)\hat{\varrho}_{\kappa_2}^2\right)\chi_{\kappa_2}(\theta)=\kappa_2\left(\frac{1}{4}-a_1^2\right)\chi_{\kappa_2}(\theta).
\end{equation}
We able to introduce the new differential operator in a simpler form of Eq. \eqref{eq28} as
\begin{equation}\label{eq29}
\hat{b}_{\kappa_2}=-C_{\kappa_2}^2(\theta)\frac{\partial^2}{\partial\theta^2}+\frac{\left(\epsilon^2-\tfrac{1}{4}\right)}{T^2_{\kappa_2}(\theta)}-\kappa_2C_{\kappa_2}^2(\theta)\hat{\varrho}^2_{\kappa_2},
\end{equation}
it can be represented by the following factorization property
\begin{equation}\label{eq34}
\hat{b}_{\kappa_2}=A_{\kappa_2,\hat{\varrho}_{\kappa_2}}^-A_{\kappa_2,\hat{\varrho}_{\kappa_2}}^++\hat{\lambda}_{\kappa_2},
\end{equation}
where
\begin{equation}\label{eq30}
\begin{split}
A_{\kappa_2,\hat{\varrho}_{\kappa_2}}^-=& C_{\kappa_2}(\theta)\frac{\partial}{\partial\theta}+\frac{\left(\epsilon+\tfrac{1}{2}\right)}{S_{\kappa_2}(\theta)}+(\hat{\varrho}_{\kappa_2}+\kappa_1)S_{\kappa_2}(\theta),\\
A_{\kappa_2,\hat{\varrho}_{\kappa_2}}^+=& -C_{\kappa_2}(\theta)\frac{\partial}{\partial\theta}+\frac{\left(\epsilon+\tfrac{1}{2}\right)}{S_{\kappa_2}(\theta)}+\hat{\varrho}_{\kappa_2}S_{\kappa_2}(\theta),
\end{split}
\end{equation}
and
\begin{equation}
\hat{\lambda}_{\kappa_2}=-\kappa_2\left(\kappa_2\epsilon+\hat{\varrho}_{\kappa_2}+\tfrac{3\kappa_2}{2}\right)\left(\kappa_2\epsilon+\hat{\varrho}_{\kappa_2}+\tfrac{\kappa_2}{2}\right).
\end{equation}
It is understood that evolution of reflective operators $R_1=\pm1$ on the zenithal wavefunction is caused to single indicator variable $a_1$, read
\begin{equation}
\ \begin{cases}
a_1^+=\frac{1}{2}-\mu_1, \quad\ \qquad \textrm{if} \quad e_1=1\\
a_1^-=\frac{1}{2}+\mu_1, \quad\qquad \ \textrm{if}  \quad e_1=-1
\end{cases},
\end{equation}
The action of the ladder operators \eqref{eq30} on the differential operator of $\hat{b}_{\kappa_2}$ is
\begin{equation}\label{eq41}
A_{\kappa_2,\hat{\varrho}_{\kappa_2}}^+\hat{b}_{\kappa_2}(\hat{\varrho}_{\kappa_2})=\hat{b}_{\kappa_2}(\hat{\varrho}_{\kappa_2}+1)A_{\kappa_2,\hat{\varrho}_{\kappa_2}}^+, \qquad A_{\kappa_2,\hat{\varrho}_{\kappa_2}}^-\hat{b}_{\kappa_2}(\hat{\varrho}_{\kappa_2}+1)=\hat{b}_{\kappa_2}(\hat{\varrho}_{\kappa_2})A_{\kappa_2,\hat{\varrho}_{\kappa_2}}^-.
\end{equation}
As a result of acting on differential operator of $\hat{b}_{\kappa_2}$, their commutator relations are as follows
\begin{equation}
[\hat{b}_{\kappa_2}(\hat{\varrho}_{\kappa_2}),A_{\kappa_2,\hat{\varrho}_{\kappa_2}}^-]=-A_{\kappa_2,\hat{\varrho}_{\kappa_2}}^-(2(a_2+\kappa_2\hat{\varrho}_{\kappa_2}+1)+\kappa_2), \qquad [\hat{b}_{\kappa_2}(\hat{\varrho}_{\kappa_2}),A_{\kappa_2,\hat{\varrho}_{\kappa_2}}^+]=(2(a_2+\kappa_2\hat{\varrho}_{\kappa_2}+1)+\kappa_2)A_{\kappa_2,\hat{\varrho}_{\kappa_2}}^+.
\end{equation}
In analogues of the ladder operators in making use of the transformation \eqref{eq32}, and removing the first-order derivative for the shift operators of $\hat{\mathcal{H}}_\kappa$, we consider the following reduced eigenfunction
\begin{equation}
R_{\kappa_1}(r)=\frac{\rho_{\kappa_1}(r)}{(C_{\kappa_1}(r))^{\mu_0}(S_{\kappa_1}(r))^{(1+\mu_1+\mu_2+\mu_3)}}, 
	\end{equation}
hence, the result of the Hamiltonian for $\hat{\mathcal{H}}_\kappa$ in Eq. \eqref{eq10} is
\begin{equation}
-\frac{1}{2}\left[\frac{\partial^2}{\partial r^2}-\frac{\hat{\varrho}_{\kappa_2}^2-\tfrac{1}{4}}{S_{\kappa_1}^2(r)}-\frac{\kappa_1\left(\mu_0(\mu_0-R_0)+\omega^2\right)}{C_{\kappa_1}^2(r)}+\kappa_1\left((1+\mu_0+\mu_1+\mu_2+\mu_3)^2+\omega^2\right)\right],
\end{equation}
where its form is equivalent to a two-parameter curved Dunkl P\"{o}schel-Teller Hamiltonian with reflection $R_0$. However, one can represent by the following factorization property
\begin{equation}\label{eq33}
\hat{\mathcal{H}}_\kappa=\hat{B}^-_{\kappa_1}\hat{B}^+_{\kappa_1}+\hat{\lambda}_{\kappa_1},
\end{equation}
where the intertwining operators here mean the same shift operators, namely
\begin{equation}\label{eq42}
\begin{split}
\hat{B}^+_{\kappa_1}&=\frac{\hbar}{\sqrt{2m}}\left(-\frac{\partial}{\partial r}-\kappa_1b_1T_{\kappa_1}(r)+\frac{b_2}{T_{\kappa_1}(r)}\right),\\
\hat{B}^-_{\kappa_1}&=\frac{\hbar}{\sqrt{2m}}\left(\frac{\partial}{\partial r}-\kappa_1b_1T_{\kappa_1}(r)+\frac{b_2}{T_{\kappa_1}(r)}\right),
\end{split}
\end{equation}
with a spectrum of
\begin{equation}
\hat{\lambda}_{\kappa_1}=\frac{\kappa_1}{2}\left[(1+\mu_0+\mu_1+\mu_2+\mu_3)^2+(b_1+b_2)^2+\omega^2\right].
\end{equation}
Here $b_1$ is the indicator variables of the reflection $R_0$ on the $\rho_{\kappa_1}(r)$
which determines as 
\begin{equation}
\begin{cases}
b_1^+=\frac{1}{2}\left(1-\sqrt{4\omega^2+(1-2\mu_0)^2}\right), \quad\ \qquad \textrm{if} \quad R_0=1\\
b_1^-=\frac{1}{2}\left(1-\sqrt{4\omega^2+(1+2\mu_0)^2}\right),\quad \qquad \ \textrm{if} \quad R_0=-1
\end{cases},
\end{equation}
and with the parameter of $b_1=\frac{1}{2}\left(\hat{\varrho}_{\kappa_2}+1\right)$.
Consequently, we can show that the shift operators $\hat{B}^\pm_{\kappa_1}$ satisfy the commutation relations
\begin{equation}
[\hat{B}^-_{\kappa_1},\hat{B}^+_{\kappa_1}]=\frac{\kappa_1b_1}{C_{\kappa_1}^2(r)}+\frac{b_2}{S_{\kappa_1}^2(r)},\qquad
[\hat{\mathcal{H}}_\kappa,\hat{B}^\pm_{\kappa_1}]=\left(\frac{\kappa_1b_1}{C_{\kappa_1}^2(r)}+\frac{b_2}{S_{\kappa_1}^2(r)}\right)\hat{B}^\pm_{\kappa_1}.
\end{equation}
Now, we can define the additional symmetry operators $\hat{\mathcal{X}}^\pm_\kappa$ which are the simplest case of the curved Hamiltonian $\hat{\mathcal{H}}_\kappa$ given in \eqref{eq10}, in the geodesic parallel coordinates by combining a pair of symmetry
operators \eqref{eq41} and \eqref{eq42} in the following way
\begin{equation}\label{eq50}
\hat{\mathcal{X}}^\pm_\kappa=(\hat{B}^\pm_{\kappa_1})(\hat{A}^\pm_{\kappa_2,\hat{\varrho}_{\kappa_2}})=\pm\hat{\mathcal{O}}'_\kappa\hat{\varrho}_{\kappa_2}+\hat{\mathcal{W}}'_\kappa,
\end{equation}
so that $[\hat{\mathcal{H}}_\kappa,\hat{\mathcal{X}}_\kappa^\pm]=0$.
The symmetries $\hat{\mathcal{W}}'_\kappa$ and $\hat{\mathcal{O}}'_\kappa$ are polynomials in the momentum operators. 
Generally, the Hamiltonian $\hat{\mathcal{H}}_\kappa$ \eqref{eq10} defines an integrable system, since it commutes with the operator $B_{\kappa_2}^\theta$ \eqref{eq17}. Therefore, the Hamiltonian $\hat{\mathcal{H}}_\kappa$ is read as a superintegrable curved Dunkl oscillator with additional symmetry operators given by \eqref{eq51} and \eqref{eq50}. Hence, the sets $(\hat{\mathcal{H}}_\kappa,B_{\kappa_2}^\theta,\hat{\mathcal{X}}^+_\kappa)$ and $(\hat{\mathcal{H}}_\kappa,B_{\kappa_2}^\theta,\hat{\mathcal{X}}^-_\kappa)$ are formed by algebraically independent operators.
\subsection{Separated Solutions in geodesic polar coordinates}\label{sec4}
In order to obtain the exact solution of Eq. \eqref{eq10}, we display that separation of the coordinate system \eqref{eq37} leads to equations of
\begin{align}
 \left(A_r^{\kappa_1}+\frac{L}{2S_{\kappa_1}^2(r)}\right)R_{\kappa_1}(r)&=\mathcal{E}_\kappa R_{\kappa_1}(r), \label{eq38}\\
 \left(B_\theta^{\kappa_2}+\frac{\ell_z}{2S_{\kappa_2}^2(\theta)}\right)\Theta_{\kappa_2}(\theta)&=L \Theta_{\kappa_2}(\theta), \label{eq39}\\
\ C_\varphi\Phi(\varphi)&=\ell_z \Phi(\varphi), \label{eq40}
\end{align}
where $L$ and $\ell_z$ are the separation constants. The solutions of \eqref{eq39} end \eqref{eq40} have been obtained in \cite{Hamiltonian} by using the parametric form of the NU method \cite{nu}. For the angular equation \eqref{eq40}, the solutions derived by the quantum number $(e_2,e_3)$ corresponding to the reflection operators $(R_2,R_3)$ are given by the eigenvalues $e_2,e_3=\pm1$ in the following simple form
\begin{equation}
\Psi_m^{(e_2,e_3)}(\varphi)=\mathcal{N}^{(e_2,e_3)}_m\left(\sin\varphi\right)^{\tfrac{1}{2}(1-e_3)}\left(\cos\varphi\right)^{\tfrac{1}{2}(1-e_2)}P^{\left(\mu_3-\tfrac{e_3}{2},\mu_2-\tfrac{e_2}{2}\right)}_m(\cos2\varphi)
\end{equation}
where $P_n^{(\alpha,\beta)}(x)$ denotes the Jacobi polynomials \cite{jacobi}and the normalization constant is
\begin{equation*}
\mathcal{N}^{(e_2,e_3)}_m=\left[\frac{(2m+\mu_2+\mu_3)\Gamma\left(m+\mu_2+\mu_3-\tfrac{e_2}{4}-\tfrac{e_3}{4}+\tfrac{1}{2}\right)\left(m+\tfrac{e_2}{4}+\tfrac{e_3}{4}-\tfrac{1}{2}\right)!}{2\Gamma\left(m+\mu_2-\tfrac{e_2}{4}+\tfrac{e_3}{4}+\tfrac{1}{2}\right)\Gamma\left(m+\mu_3+\tfrac{e_2}{4}-\tfrac{e_3}{4}+\tfrac{1}{2}\right)}\right]^{\tfrac{1}{2}}.
\end{equation*}
 This approach causes the separation constant
$\ell_z=4m(m+\mu_2+\mu_3)$. The zenital solutions $\Theta_{\ell,\kappa_2}^{(e_1)}(\theta)$ with the eigenvalues $e_1=\pm1$ of the reflection operator $R_1$ are determined by
\begin{equation}
\Theta_{\ell,\kappa_2}^{(e_1)}(\theta)=\mathcal{N}_\ell^{(e_1)} \left(S_{\kappa_2}(\theta)\right)^{2m}\left(C_{\kappa_2}(\theta)\right)^{\tfrac{1}{2}(1-e_1)}P^{\left(2m+\mu_2+\mu_3,\mu_1-\tfrac{e_1}{2}\right)}_\ell(C_{\kappa_2}(2\theta)),
\end{equation}
with the normalization constant
\begin{equation*}
\mathcal{N}_\ell^{(e_1)}=\left[\frac{(2\ell+2m+\mu_1+\mu_2+\mu_3+\tfrac{1}{2})\Gamma\left(\ell+2m+\mu_1+\mu_2+\mu_3-\tfrac{e_1}{4}+\tfrac{3}{4}\right)\left(\ell+\tfrac{e_1}{4}-\tfrac{1}{4}\right)!}{\Gamma\left(\ell+2m+\mu_2+\mu_3+\tfrac{e_1}{4}+\tfrac{3}{4}\right)\Gamma\left(\ell+\mu_1-\tfrac{e_1}{4}+\tfrac{3}{4}\right)}\right]^{\tfrac{1}{2}}.
\end{equation*}
where the value of the separation constant is
$L=4\kappa_2(\ell+m)\left(\ell+m+\mu_1+\mu_2+\mu_3+\tfrac{1}{2}\right)$. On the other hand, the radial solutions with the quantum number $e_0=\pm 1$ of the reflection operator $R_0$ are given by
\begin{equation}
R_{n,\kappa_1}^{(e_0)}(r)=\mathcal{N}_n^{(e_0)} \left(S_{\kappa_1}(r)\right)^{2(\ell+m)}\left(C_{\kappa_1}(r)\right)^{\tfrac{1}{2}(1-e_0)+\omega'-\mu_0}P^{\left(2(\ell+m)+\mu_1+\mu_2+\mu_3+\tfrac{1}{2},\omega'-\tfrac{e_0}{2}\right)}_n(C_{\kappa_1}(2r)),
\end{equation}
the
normalization constant has the expression
\begin{equation*}
\mathcal{N}_n^{(e_0)}=\left[\frac{(2n+2\ell+2m+\mu_0+\mu_1+\mu_2+\mu_3+1)\Gamma\left(n+2\ell+2m+\mu_1+\mu_2+\mu_3-\tfrac{e_0}{4}+\tfrac{5}{4}\right)\left(n+\tfrac{e_0}{4}-\tfrac{1}{4}\right)!}{\Gamma\left(\ell+2m+\mu_2+\mu_3+\tfrac{e_1}{4}+\tfrac{5}{4}\right)\Gamma\left(n+\mu_0-\tfrac{e_0}{4}+\tfrac{3}{4}\right)}\right]^{\tfrac{1}{2}}.
\end{equation*}
So, the total energy of the separated wavefunctions 
$
\Psi_{\kappa, n,\ell,m}^{(e_0,e_1,e_2,e_3)}(r,\theta,\varphi)=R_{n,\kappa_1}^{(e_0)}(r)\Theta_{\ell,\kappa_2}^{(e_1)}(\theta)\Phi_m^{(e_2,e_3)}(\varphi)
$
of the three-dimensional Dunkl oscillator in the geodesic polar coordinates reads
\begin{equation}
\mathcal{E}_\kappa=\frac{\kappa_1}{2}\left[\left(\lambda+2(n+\ell+m)\right)^2+\omega'-\lambda^2-3\mu_0+2(\lambda-\omega'-1)(\omega'-\mu_0)\right]
\end{equation}
in which
$\lambda=\mu_1+\mu_2+\mu_3+\omega'+1$.
It is directly seen that the wave functions obey
the orthogonality relations
\begin{align}
\begin{split}
&\int^\infty_{0}\int^{\pi}_{0}\int^{2\pi}_{0}\Psi_{\kappa, n,\ell,m}^{(e_0,e_1,e_2,e_3)}(r,\theta,\varphi)\Psi_{\kappa, n',\ell',m'}^{(e'_0,e'_1,e'_2,e'_3)}(r,\theta,\varphi)|C_\kappa(r)|^{2\mu_0}|S_{\kappa_1}(r)C_{\kappa_2}(\theta)|^{2\mu_1}|S_{\kappa_1}(r)S_{\kappa_2}(\theta)\cos\varphi|^{2\mu_2}\\
&|S_{\kappa_1}(r)S_{\kappa_2}(\theta)\sin\varphi|^{2\mu_3}S_{\kappa_1}^2(r)S_{\kappa_2}(\theta)\mathrm{d}r\mathrm{d}\theta\mathrm{d}\varphi=\delta_{n,n'}\delta_{\ell,\ell'}\delta_{m,m'}\delta_{e_0,e_0'}\delta_{e_1,e_1'}\delta_{e_2,e_2'}\delta_{e_3,e_3'}.
\end{split}
\end{align}
On the other hand, these wavefunctions appear simultaneously with the reflective operators $R_i$ with eigenvalues $e_i = \pm1$ for $i = 0,1, 2, 3$. This reflects the statement that when $e_0 = 1$, the quantum number $n$ takes non-negative integer values and when $e_0 = -1$, the number $n$ takes positive half-integer values. The same is true for quantum number $\ell$, that is, by having $e_1 = 1$ the quantum
number $\ell$ is a non-negative integer and when $e_1 = -1$, $\ell$ is a positive half-integer. Similarly, when $e_2e_3 = 1$, the quantum
number $m$ becomes a non-negative integer and when $e_2e_3 = -1$, $m$ becomes a positive half-integer.
\section{Conclusions}\label{sec5}
In this letter, we studied the generalization of the Dunkl oscillator model in the plane to the deformation space $\mathbb{S}^3_{[\kappa_1]\kappa_2}$ by applying reflections and in making use of the Jordan-Schwinger representation. We have explicitly shown presented superintegrability for this model and have seen obtained constants of motion satisfy the symmetry algebra $jsd_{\kappa_1\kappa_2}(4)$, called Jordan-Schwinger-Dunkl algebra.
In addition, we have shown that under the addition of some centrifugal terms, due to the use of gauge transformation on one-dimensional Hamiltonians to the potential term, and the curved analogues of these resulted Hamiltonians can also be considered as superintegrable.
In order to get other symmetries concerning superintegrability, we applied a factorization approach.
Moreover, we obtained the solutions of the energy spectrum and wave functions in the geodesic polar coordinates for the present model.


\begin{thebibliography}{99}
\bibitem{Hamiltonian}
V.X. Genest, L. Vinet, A. Zhedanov, J. Phys.: Conf. Ser.  512 (2014) 012010. 
\bibitem{derivative}
C.F. Dunkl, Trans. Amer. Math. Soc. 311 (1989) 167-183.
\bibitem{operator}
D. Ojeda-Guill\'{e}n, R.D. Mota, M. Salazar-Ram\'{\i}rez, V.D. Granados, Mod. Phys. Lett. A, 35(31) (2020) 2050255.
\bibitem{Hamiltonian1}
V.X. Genest, M.E. Ismail, L. Vinet, A. Zhedanov, J. Phys. A: Math. Theor. 46 (2013) 145201.
\bibitem{Hamiltonian2}
V.X. Genest,  M.E. Ismail, L. Vinet, A. Zhedanov, Comm. Math. Phys. 329 (2014) 999-1029.
\bibitem{cayley}
N.A. Gromov,  V.I. Man'ko, J. math. phys. 31 (1990) 1047-1053.
\bibitem{coulomb}
V.X. Genest, A. Lapointe, L. Vinet, Phys. Lett. A, 379 (2015) 923-927.
\bibitem{coulomb1}
S. Ghazouani, I. Sboui, M.A. Amdouni, M.B.E.H. Rhouma, J. Phys. A: Math. Theor. 52 (2019) 225202.
\bibitem{coulomb2}
S. Ghazouani, and S. Insaf, J. Phys. A: Math. Theor. 53 (2019) 035202.
\bibitem{coulomb3}
S. Ghazouani, Ana. Math. Phys. 11 (2021) 1-99.
\bibitem{wilson}
V.X. Genest, L. Vinet, A. Zhedanov, J. Phys. A: Math. Theor. 46 (2013) 325201.
\bibitem{wilson1}
M. Salazar-Ramırez, D. Ojeda-Guill\'{e}n, R.D. Mota, and V.D. Granados, Eur. Phys. J. Plus, 132 (2017) 1-8.
\bibitem{wilson2}
M. Salazar-Ram\'{\i}rez, D. Ojeda-Guill\'{e}én, R.D. Mota, V.D. Granados, Mod. Phys. Lett. A, 33 (2018) 1850112.
\bibitem{sphere}
V.X. Genest, L. Vinet, and A. Zhedanov, Comm. Math. Phys. 336 (2015) 243-259.
\bibitem{sphere1}
V.X. Genest, L. Vinet, and A. Zhedanov, J. Phys. A: Math. Theor. 47 (2014) 205202.
\bibitem{sphere2}
H. De Bie, V.X. Genest, J.M. Lemay, L. Vinet, Acta Poly. 56 (3) (2016) 166-172. 
\bibitem{sphere3}
H. De Bie, V.X. Genest, J.M. Lemay, L. Vinet, J. Phys. A: Math. Theor. 50(19) (2017) 195202.
\bibitem{factor}
\c{S}. Kuru, and J. Negro, Ann. Phys. 323(2) (2008), 413.
\bibitem{factor1}
J. A. Calzada, \c{S}. Kuru, and J. Negro, Eur. Phys. J. Plus, 129(8) (2014), 164.
\bibitem{factor2}	
J. A. Calzada, \c{S}. Kuru, J. Negro, and M. A. Del Olmo, Ann. Phys. 327(3) (2012), 808.
\bibitem{change1}
M.F. Ra\~{n}ada, M. Santander, J. Math. Phys. 40(10) (1999) 5026-5057.
\bibitem{change2}
F.J. Herranz, and M. Santander, J. Phys. A: Math. Gen. 35(31) (2002) 6601.
\bibitem{am1}
A. Najafizade, H. Panahi, H. Hassanabadi, Phys. A: Stat. Mech. Appl. 543 (2020) 123414.
\bibitem{am2}
A. Najafizade, H. Panahi, Phys. A: Stat. Mech. Appl. 573 (2021) 125935.
\bibitem{proper}
F.J. Herranz, R. Ortega, and M. Santander, J. Phys. A: Math. Gen. 33 (24) (2000) 4525.
\bibitem{phiggs}
J.F. Carinena, M.F. Ranada, M. Santander, M. Senthilvelan, Nonlinearity, 17 (5) (2004) 1941.
\bibitem{sl}
S. Tsujimoto, L. Vinet, A. Zhedanov, Sym. Integ. Geo. Meth. Appl. 7 (2011) 093.
\bibitem{gauge}
V.X. Genest, J.M. Lemay, L. Vinet, A. Zhedanov, J. Phys. A: Math. Theor. 46 (50) (2013) 505204.
\bibitem{LB1}
E.G. Kalnins, J.M. Kress, W. Miller Jr., P. Winternitz, J. Math. Phys. 44 (12) (2003) 5811.
\bibitem{LB2}
\'{A}. Ballesteros, A. Enciso, F.J. Herranz, O. Ragnisco, D. Riglioni, Ann. Phys. 326 (8) (2011) 2053.
\bibitem{sw}
C. Grosche, G.S. Pogosyan, and A.N. Sissakian, Fortschritte der Physik/Prog. Phys. 43 (6) (1995) 523-563.
\bibitem{sw1}
E.G. Kalnins, W. Miller Jr, and G.S. Pogosyan, J. Math. Phys. 38 (10) (1997) 5416-5433.
\bibitem{sw2}
F.J. Herranz, \'{A}. Ballesteros, M. Santander, T. Sanz-Gil, CRM Proc. and Lecture Notes, 37 (2004) 75-89.
\bibitem{sw3}
\'{A}. Ballesteros, F.J. Herranz, M. Santander, T. Sanz-Gil, J. Phys. A: Math. Gen. 36 (7) (2003) 93.
\bibitem{sw4}
F.J. Herranz, and \'{A}. Ballesteros, Sym. Integ. Geo. Meth. Appl. 2 (2006) 010.
\bibitem{aniso1}
\'{A}. Ballesteros, F.J. Herranz, F. Musso, Nonlinearity, 26(4) (2013) 971.
\bibitem{aniso2}
\'{A}. Ballesteros, A. Blasco, F.J. Herranz, F. Musso, J. Phys. A: Math. Theor. 47 (34) (2014) 345204.
\bibitem{aniso3}
\'{A}. Ballesteros, F.J. Herranz, \c{S}. Kuru, J. Negro, Ann. Phys. 373 (2016) 399-423.
\bibitem{higgs}
P.W. Higgs, J. Phys. A: Math. Gen. 12 (3) (1979) 309.
\bibitem{higgs1}
H.I. Leemon, J. Phys. A: Math. Gen. 12 (4) (1979) 489.
\bibitem{shi}
S.H. Dong, Factorization method in quantum mechanics, Vol. 150, Springer Science and Business Media, 2007.
\bibitem{poschl}
\c{S}. Kuru, and J. Negro, Ann. Phys. 324(12) 2548 (2009).
\bibitem{poschl2}
G. F. Torres del Castillo, and T. Tepper Garc\'{\i}a, Rev. Mex. F\'{\i}s. 48(1) (2002) 16.
\bibitem{nu}
C. Tezcan, R. Sever, Int. J. Theor. Phys. 48 (2) (2009) 337-350.
\bibitem{jacobi}
R. Koekoek, P.A. Lesky, R.F. Swarttouw, Hypergeometric orthogonal polynomials and their q-analogues, 1st edition, Springer,  2010.
\end{thebibliography}
\end{document}